\documentclass[apj]{emulateapj}

\usepackage{natbib}
\usepackage{graphicx}
\usepackage{bm}
\usepackage{amsmath}
\usepackage{amsfonts}
\usepackage{amssymb}

\pdfoutput=1

\def\gtwid{\mathrel{\raise.3ex\hbox{$>$\kern-.75em\lower1ex\hbox{$\sim$}}}}
\def\ltwid{\mathrel{\raise.3ex\hbox{$<$\kern-.75em\lower1ex\hbox{$\sim$}}}}
\def\\{\hfil\break}

\newcommand{\R}{\textbf{R}\ }
\newcommand{\Mtot}{$M_{\text{tot}}$~}
\newcommand\T{\rule{0pt}{2.6ex}}       

\def\sun{\hbox{$\odot$}}

\def\lesssim{\mathrel{\hbox{\rlap{\hbox{\lower4pt\hbox{$\sim$}}}\hbox{$<$}}}}
\def\gtrsim{\mathrel{\hbox{\rlap{\hbox{\lower4pt\hbox{$\sim$}}}\hbox{$>$}}}}

\newcommand{\mamo}[1]{\mbox{$#1$}}
\newcommand{\unit}[1]{\ifmmode \:\mbox{\rm #1}\else \mbox{#1}\fi}
\newcommand{\mone}{\mamo{^{-1}}}

\newcommand{\kms}{\unit{km~s\mone}}

\newcommand{\kpc}{\unit{kpc}}

\newcommand{\msun}{\mamo{M_{\sun}}}

\begin{document}

\title{Estimating the Galactic Mass Profile in the Presence of Incomplete Data}
\author{Gwendolyn M. Eadie\altaffilmark{1}}
\author{William E. Harris\altaffilmark{1}}
\author{Lawrence M. Widrow\altaffilmark{2}}
\altaffiltext{1}{Dept.\ of Physics and Astronomy, McMaster University, Hamilton, ON L8S 4M1, Canada.}
\altaffiltext{2}{Dept.\ of Physics, Engineering Physics \& Astronomy, Queen's University, Kingston, ON K7L 3N6, Canada.}

\shorttitle{}
\shortauthors{Eadie, Harris, and Widrow}

\begin{abstract}
A powerful method to measure the mass profile of a galaxy is through the velocities of tracer particles distributed through its halo. Transforming this kind of data accurately to a mass profile $M(r)$, however, is not a trivial problem. In particular, limited or incomplete data may substantially affect the analysis. In this paper we develop a Bayesian method to deal with incomplete data effectively; we have a hybrid-Gibbs sampler that treats the unknown velocity components of tracers as parameters in the model. We explore the effectiveness of our model using simulated data, and then apply our method to the Milky Way using velocity and position data from globular clusters and dwarf galaxies. We find that in general, missing velocity components have little effect on the total mass estimate. However, the results are quite sensitive to the outer globular cluster Pal 3. Using a basic Hernquist model with an isotropic velocity dispersion, we obtain credible regions for the cumulative mass profile $M(r)$ of the Milky Way, and provide estimates for the model parameters with 95\% Bayesian credible intervals. The mass contained within 260kpc is $1.37\times10^{12}$\msun, with a 95\% credible interval of $(1.27,1.51)\times10^{12}$\msun. The Hernquist parameters for the total mass and scale radius are $1.55^{+0.18}_{-0.13}\times10^{12}$\msun~ and $16.9^{+4.8}_{-4.1}$~kpc, where the uncertainties span the 95\% credible intervals. The code we developed for this work, Galactic Mass Estimator (GME), will be available as an open source package in the R Project for Statistical Computing.

\end{abstract}


\maketitle

\label{firstpage}

\section{Introduction}

Almost every galaxy in the universe is assumed to reside in a massive, dark matter halo that can extend far beyond the visible components of the galaxy. Standard methods to determine the mass distribution within visible portions of galaxies are based on rotation curves or velocity dispersion profiles. The former is applicable to spiral galaxies, while the latter method is used mainly for elliptical galaxies or disk galaxy bulges. At large radii, where the mass distribution is presumably dominated by dark matter, one can use observations of kinematic tracers to learn about a galaxy's mass profile.

The Milky Way (MW) has many distant satellites, such as globular clusters, halo stars, planetary nebulae, and dwarf galaxies. The kinematic properties of these satellites can be used to learn about the gravitational potential of the whole system, and thus the Galaxy's mass profile out to large radii.

One way to use the kinematic data of tracers to estimate the mass distribution is via a \emph{mass estimator}, a method first suggested by \cite{hartwick1978} and a method that avoids using an explicit model. The method has endured partly because it uses only the line of sight velocities and positions of the tracers. It is useful when no proper motions are available and conversion to a Galactocentric refrence frame is impossible. Nowadays, however, many MW tracers have proper motion measurements, and more continue to become available. Thus, it would be beneficial to have a method which can incorporate both complete data (i.e.~tracers with 3-dimensional space motions in the Galactocentric frame) and incomplete data (i.e.~tracers with only line of sight velocities). In this paper we introduce such a method of mass estimation.

The method used here is based on the phase-space distribution function (DF), which is a probability distribution for a satellite in terms of its position $\bm{r}$ and velocity $\bm{v}$ in the Galactocentric reference frame. Using the DF for a model and a Bayesian approach to the analysis, we obtain probability distributions for the model parameters (and thus the mass). The  method was first suggested by \cite{little1987}, who showed how to use the DF and a Bayesian approach to estimate the mass specifically for the Milky Way. Since then, other studies have used the Bayesian approach or Maximum Likelihood methods for Galactic mass estimation \citep[e.g.][]{kulessa1992, kochanek1996, wilkinsonevans1999, widrow2008}. Deriving analytic and physically relevant DFs has been explored by \cite{hernquist1990}, \cite{1991cuddeford}, \cite{ciotti1996}, \cite{widrow2000}, and \cite{evans2014}, to name a few.

A DF's derivation and final form is, by default, in the Galactocentric reference frame, but previous studies have re-written DFs in terms of the line-of-sight velocity component only, in order to incorporate incomplete data \citep[e.g.][]{wilkinsonevans1999}. This does not take full advantage of the complete data that is available, which is an issue when the method may be susceptible to large uncertainties due to small sample size \citep[as discussed by][]{zaritsky1989}. Furthermore, rewriting the DF in terms of the line-of-sight velocity can be mathematically difficult. In our study, we introduce a generalized approach via the Bayesian framework, whereby it is easy to incorporate complete and incomplete data simultaneously, and also unnecessary to rewrite the DF in terms of the line-of-sight velocity.

The purpose of this paper is to lay out the method fully, and set the groundwork for future studies with a range of DFs and datasets. We also test the method with simulated data, and do some preliminary analysis with the method as it applies to the Milky Way.

The outline of this paper is as follows. Section~\ref{sec:df} briefly describes the theoretical background of DFs, and Section~\ref{sec:models} introduces two models that already have analytic DFs, which we use in this work. Next, in Section~\ref{sec:Bayes}, we review how Bayes' Theorem can be used with the DFs to obtain parameter estimates of the model. Sections~\ref{sec:newmethods}, ~\ref{sec:gibbs}, and~\ref{sec:priors} introduce the method for incorporating both complete and incomplete data via a hybrid-Gibbs Sampler, and Section~\ref{sec:convergence} discusses the techniques used to assess convergence of the Markov chains.

We first apply and test our method on simulated data (described in Section~\ref{sec:sims}) and then apply the method to some kinematic data of satellites orbiting the Milky Way (described in Section~\ref{sec:data}). The results of these analyses are presented and discussed in Sections~\ref{sec:results} and~\ref{sec:discussion} respectively. Many future prospects are also discussed in  Section~\ref{sec:discussion}.

\section{Background}
This section provides background information and notation about distribution functions, the models we use in our analysis, and the methods of Bayesian inference as they apply to the current problem. Additional details and discussion can be found in \cite{eadieMSc2013}. 

\subsection{Distribution Function}\label{sec:df}

The Distribution Function (DF), $f(\bm{r},\bm{v})$, is a probability density function that gives the probability of finding a particle with a position $\textbf{r}$ and velocity $\textbf{v}$ within a phase-space volume element $d^3\textbf{r}d^3\textbf{v}$ \citep{binney2008}. Like any probability density, the DF integrates to one:
\begin{equation}\label{eq:fint}
	\int{f(\textbf{r},\textbf{v})d^3\textbf{r}d^3\textbf{v}} = 1.
\end{equation}
Eq.~\ref{eq:fint} is often renormalized so that the DF integrates to a quantity of interest, such as the total mass $M_{\text{tot}}$:
\begin{equation}\label{eq:fintM}
	\int{f(\textbf{r},\textbf{v})d^3\textbf{r}d^3\textbf{v}} = M_{\text{tot}}.
\end{equation}
However, in a Bayesian framework the DF as defined in eq.~\ref{eq:fint} is used, and thus the left-hand-side of eq.~\ref{eq:fintM} is divided by $M_{\text{tot}}$. Thus, it is important that models have a finite mass in a Bayesian analysis--- if the mass is infinite, then the DF is not a proper probability distribution.

A DF can be specified by use of Jeans' Theorem, which states that any solution of the time-independent collisionless Boltzmann equation is a function of the phase-space coordinates ($\bm{r},\bm{v}$) only. In a time-independent system, the Hamiltonian $H=\frac{v^2}{2}+\Phi(r)$ is always an integral of motion, and if the system is also spherical then the magnitude of the angular momentum, $L$, is an integral of motion too. Therefore, any non-negative function $f(H)$ or $f(H,L)$  will be a solution to the time-independent collisionless Boltzmann equation, and thus a DF for the system. Whether or not $f$ is a function of $H$ or both $H$ and $L$ determines the velocity dispersion of the system; $f(H)$ corresponds to an isotropic system, and $f(H,L)$ corresponds to an anisotropic system.

In practice, a DF corresponding to an isotropic, spherical, self-consistent system is usually written in terms of $\mathcal{E}$, the relative energy per unit mass, defined as
\begin{equation}
 \mathcal{E}=-\frac{v^{2}}{2} + \Psi(r)
	\label{eq:scriptE}
\end{equation}
where $v$ is the speed of a particle at a distance $r$ from the center of the system, and $\Psi(r)$ is the relative gravitational potential of the system at $r$, as defined in \cite{binney2008}. Particles with $\mathcal{E}\leq 0$ are unbound and require $f=0$. If the system has an anisotropic velocity dispersion, then the DF is written as a function of both $\mathcal{E}$ and the angular momentum $L$.

\subsection{Models}\label{sec:models}
Models with analytic DFs are preferable to empirical distribution functions in theoretical analyses because they allow for easy sampling of the distribution, and also save computation time by avoiding numerical integration. Finding a DF that models a realistic galaxy is a difficult task, however, because galaxies are often composed of multiple subsystems such as a bulge, a stellar halo, a dark matter halo, and possibly a disk. Finding a single phase-space distribution function that is self-consistent, analytic, and that describes the intricate features of a galaxy is very challenging. 

An empirical luminosity profile that has been successful in fitting the surface brightness profiles of elliptical galaxies and bulges is the \cite{devaucouleurs1948}  $R^{1/4}$ profile. A generalization of the $R^{1/4}$ profile is $R^{1/n}$, which was introduced by \cite{sersic1968}. Due to the success of $R^{1/4}$, theorists have tried to develop distribution functions that can reproduce the profile. The analytic models introduced by \cite{jaffe1983} and \cite{hernquist1990} fit the $R^{1/4}$ type galaxies well for most radii.

In this work, primarily for the purpose of testing the method, we use the Hernquist and Jaffe models because of their analytic simplicity and their ubiquity. The Hernquist model also has the benefit of having more than one analytic DF--- one has an isotropic velocity dispersion, and there are a few that are anisotropic. Furthermore, both the Hernquist and Jaffe models are self-consistent and have a finite total mass, making consistent numerical computations feasible. For these reasons, we use the Hernquist-style models to lay out the methodology of our Bayesian approach and the derivation of mass profile credible regions. In future work, we will extend the method to include models with non-analytic DFs and non-finite mass distributions, such as the \cite{nfw1996} (NFW) model.

\cite{hernquist1990} introduced a halo model that is a self-consistent, analytic potential-density pair. With $G\equiv1$, the gravitational potential of the Hernquist model is
\begin{equation}\label{eq:potential}
	\Phi(r)=-\frac{M_{\text{tot}}}{r+a}
\end{equation}
and the mass density profile is
\begin{equation}\label{eq:rhor}
 \rho(r)=\frac{aM_{tot}}{2\pi r\left(r+a\right)^{3}}
\end{equation}
where $M_{\text{tot}}$ is the total mass of the system, and $a$ is the scale radius. Integrating over a sphere, the Hernquist cumulative mass profile is then,
\begin{equation}\label{eq:M(r)}
 M(r) = M_{\text{tot}}\frac{r^{2}}{(r+a)^2}
\end{equation}

\cite{hernquist1990} provides two DFs for their model that are written in terms of elementary functions: one for an isotropic velocity dispersion, and a second for an anisotropic velocity dispersion of the \cite{osipkov1979} and \cite{merritt1985} type (hereafter OM-type). The OM-type allows the anisotropy to vary as a function of $r$, and includes a constant parameter called the anisotropy radius $r_a$:
\begin{equation}\label{eq:OM}
\beta(r) = \frac{r^2}{r^2 + r_a^2}
\end{equation}
The parameter $r_a$ controls the degree of radial anisotropy in the system at different radii. As $r_a\rightarrow\infty$, $\beta(r)\rightarrow0$ (completely isotropic).
 
The third Hernquist DF used in this research has a constant anisotropy $\beta=0.5$, which was derived by \cite{evans2006}. We consider this model because recent research by \cite{2013deason} showed that blue horizontal branch (BHB) stars have a radially biased velocity anisotropy of $~0.5$ between $16$ and $28$kpc, suggesting that $\beta$ may be approximately constant for most of the stellar halo.

We also consider the isotropic \cite{jaffe1983} model in this work, which has mass profile and potential 
\begin{equation}\label{eq:JaffeM}
 M_{_\text{J}}(r) = M_{\text{tot}}\frac{\ln\left(1 + a_{_\text{J}}/r\right)}{a_{_\text{J}}}
\end{equation}
\begin{equation}
 \Phi_{_\text{J}}(r)=-\frac{M_{\text{tot}}}{4\pi a_{_\text{J}} r^2\left(1+r/a_{_\text{J}}\right)^2}
\end{equation}
where $a_{\text{J}}$ is the scale radius.

Overall, we fit the Milky Way data to four different models: three Hernquist models with different anisotropies (isotropic, OM-type, constant anisotropy ($\beta=0.5$), and the isotropic Jaffe model. 

\subsection{Bayes Theorem and Parameter Estimation}\label{sec:Bayes}

Bayes' Theorem is named after Thomas Bayes (1701-1761) and was introduced posthumously by Richard Price \citep{bayes1763}. Using the rules of conditional probabilities, Bayes showed that the conditional probability $p(A|B)$ is,
\begin{equation}\label{eq:simpleBayes}
	p(A|B) = \frac{p(B|A)p(A)}{p(B)}
\end{equation}
now known as Bayes' Theorem.

Bayesian inference involves using eq.~\ref{eq:simpleBayes} in data analysis to obtain probability distributions about model parameters. Bayesian inference returns a probability distribution for parameters given the data and a prior distribution on the parameters.

When Bayes' Theorem is used for Bayesian inference, it is re-written in terms of the vector of model parameters $\bm{\theta}$ and the data $y$, and is sometimes referred to as Bayes' rule. The Bayesian posterior probability for $\bm{\theta}$, given some data $y$, is then
\begin{equation}
 p\left(\bm{\theta}|y\right)=\frac{p\left(y|\bm{\theta}\right)p\left(\bm{\theta}\right)}{p\left(y\right)}
\label{eq:BayesTheorem}
\end{equation}
where $p\left(y|\bm{\theta}\right)$ is called the \emph{likelihood}, $p\left(\bm{\theta}\right)$ is the \emph{prior probability} on the parameters, and $p\left(y\right)$ is the \emph{marginal probability} of the data. Because the marginal probability does not depend on $\bm{\theta}$, and with fixed $y$ can be considered a constant, it is common practice to sample the unnormalized posterior probability,
\begin{equation}\label{eq:BayesRule}
 p\left(\bm{\theta}|y\right)\propto p\left(y|\bm{\theta}\right)p\left(\bm{\theta}\right).
\end{equation}

The pioneering work by \cite{bahcalltremaine1981} showed that the DF determines the likelihood $p(y|\bm{\theta})$. For example, the probability of the first satellite in our data set having $r_1$ and $\bm{v}_1$, given the model parameters, is $f(r_1,\bm{v}_1|\bm{\theta})$. Assuming all $n$ satellites in the data set are independent, the probability of their corresponding positions and velocities is the product of the DFs, and thus the likelihood is
\begin{equation}\label{eq:likelihood}
p(y|\bm{\theta}) = \prod_{i=1}^nf(r_i,\bm{v}_i|\bm{\theta}).
\end{equation}
Therefore, eq~\ref{eq:BayesRule} becomes
\begin{equation}
 p\left(\bm{\theta}|y\right)\propto \prod_{i=1}^nf(r_i,\bm{v}_i|\bm{\theta})p\left(\bm{\theta}\right).
\label{eq:proportionalBayes}
\end{equation}

Sampling eq.~\ref{eq:proportionalBayes} is usually done via a Markov Chain Monte Carlo (MCMC) method, which creates a Markov chain--- a sequence of random variables, $\bm{\theta}^t$, where $t=1,2,3...$ represents the position in the chain. Every random variable in the chain depends only on the variable before it, $\bm{\theta}^{t-1}$ \citep{gelman2003}. When MCMC algorithms are used to sample a Bayesian posterior density, then by construction, the Markov chain is a collection of parameter vectors that have the same stationary distribution as the posterior (eq.~\ref{eq:BayesTheorem}). 

We apply the Metropolis algorithm  \citep{metropolis1949,metropolis1953} to sample eq.~\ref{eq:proportionalBayes}. The Metropolis algorithm is iterative and creates a Markov chain whose stationary distribution is proportional to the Bayesian posterior probability in question. The Markov chains in this work are constructed as follows \citep{gelman2003}:

\begin{enumerate}\label{list:rt}
 \item Draw a trial value $\bm{\theta^{*}}$ from a symmetric proposal distribution

 \item Calculate $d=\frac{p\left(\bm{\theta^{*}}|y\right)}{p\left(\bm{\theta}^{t-1}|y\right)}$

 \item If $d>1$, then accept $\bm{\theta^{*}}$ as $\bm{\theta}^{t}$
 \begin{enumerate}

    \item set $\bm{\theta}^{t}=\bm{\theta^{*}}$

    \item return to step 1

 \end{enumerate}

 \item If instead $d<1$, then accept $\bm{\theta^{*}}$ with probability $d$
 \begin{enumerate}
    \item draw a random number $z$ from the uniform distribution $U\left(0,1\right)$

    \item if $d>z$ then accept $\bm{\theta^{*}}$ as in step 3, and return to step 1

    \item if $d<z$ then reject $\bm{\theta^{*}}$
    \begin{enumerate}

        \item set $\bm{\theta}^{t}=\bm{\theta}^{t-1}$

        \item return to step 1
    \end{enumerate}
 \end{enumerate}

\end{enumerate}

Accepting $\bm{\theta^{*}}$ only when $d>z$ ensures that the $\bm{\theta}$ values are accepted with probability proportional to the posterior, provided that the chain has converged to the target distribution.  The above process is repeated N times, resulting in a Markov chain with N values of $\bm{\theta}$ which represents samples from the posterior. 

Because a Bayesian analysis leads to distributions for parameters, the results are arguably easier to interpret. The Markov chains can be used to acquire estimates, uncertainties, and probabilities pertaining to model parameters. Furthermore, the uncertainties can be carried forward to subsequent modeling and analysis.

A Bayesian analysis can also easily include nuisance parameters--- parameters in the model whose values are unknown but not necessarily of interest to the researcher. This feature turns out to be useful in our current problem of galaxy mass estimation; sometimes we do not know the tangential velocity component of a satellite object, but we can treat that unknown component as a nuisance parameter in the model.

\subsection{Nuisance ($v_t$) Parameters}\label{sec:newmethods}

In a Galactocentric coordinate system, the total speed of a satellite in orbit around the Galaxy can be written
\begin{equation}
	v = \sqrt{v_r^2 + v_t^2}
	\label{eq:speed}
\end{equation}
where $v_r$ and $v_t$ are the radial and tangential components respectively. In turn,
\begin{equation}
v_{t}^2 = v_{\phi}^2 + v_{\theta}^2.
\end{equation}

The total speed of a satellite is needed for the DF $f(r,v|\bm{\theta})$ in the likelihood of Bayes' theorem. However, proper motion measurements are not available for all satellites. In many cases, distant tracers of the MW have only line-of-sight velocity measurements with respect to our position in the Galaxy. We want to use this satellite information in our analyses, but without a proper motion, the line-of-sight velocity in the local standard of rest frame does not give us $v_r$ or $v_t$ in Galactocentric coordinates. For very distant objects, the line-of-sight velocity is approximately $v_{\text{los}} \approx v_r$, since the angle created by the location of the Sun, the satellite, and the center of the galaxy is quite small. However, we still have no value for $v_t$. If we treat these unknown $v_t$'s as nuisance parameters in the model, then Bayes' rule reads,
\begin{equation}
 p( \bm{\theta}|y ) \propto \prod_{i=1}^n f(r_i,v_{r,i})|\bm{\theta},v_{t,i}) p(\bm{\theta})p(v_{t,i})
\end{equation}
where $p(v_{t,i})$ is the prior probability on the tangential velocity of the $i$th satellite, to be discussed in section~\ref{sec:priors}. 

\subsection{The Gibbs Sampler}\label{sec:gibbs}

When nuisance ($v_t$) parameters are present, we use a Gibbs sampler, which was first introduced by \cite{geman1984} in the area of image processing, and then adapted to iterative simulations in the study of statistics by \cite{tanner1987}. \cite{gelfand1990} then showed how to apply it to Bayesian inference. Since then, the Gibbs sampler has been applied to many problems \cite[and references therein]{gelman2003}.

The Gibbs sampler is sometimes called alternating conditional sampling, and can be very useful in multi-dimensional problems where $\bm{\theta}=(\theta_1,...,\theta_n)$  \citep{gelman2003}. The Gibbs algorithm samples each of the parameters $(\theta_1,...,\theta_n)$ one at a time, based on the current value of all of the other parameters and the conditional probability given those parameters. 

Consider $\theta_i$, the $i$th parameter in a model with $n$ parameters, and let $\theta_{-i}$ represent all of the other parameters. Next, let $t$ be the $t^{\text{th}}$ iteration of the chain--- the chain that will be a sample of the posterior distribution $p(\bm{\theta}|y)$.  In the Gibbs sampler, each $\theta_i$ is sampled one at a time based on its conditional probability given the current values of all of the other parameters,
\begin{equation}\label{eq:conddist}
p(\theta_i|\theta^{t-1}_{-i},y)
\end{equation}
where $\theta^{t-1}_{-i}$ represents the other parameters at their current value,
\begin{equation*}
	\theta^{t-1}_{-i}=(\theta^{t}_1,...,\theta^{t}_{i-1},\theta^{t-1}_{i+1},...,\theta^{t-1}_n).
\end{equation*}

In our work, we do not directly sample the conditional distributions (eq.~\ref{eq:conddist}) because they are usually not available. Instead, we use a Metropolis step to update the conditional distributions. Thus, we employ a hybrid-Gibbs sampler: we use a symmetric proposal distribution, so that the accept/reject condition of the trial parameter ${\theta^{*}_i}$ follows the same algorithm as that described in section~\ref{sec:Bayes}, except $d$ is now
\begin{equation*}
	d = \frac{p\left(\theta^{*}_i|\theta^{t-1}_{-i},y\right)}{p\left(\theta^{t-1}_i|\theta^{t-1}_{-i},y\right)}
\end{equation*}
\citep{gelman2003}.

In the problem at hand, the hybrid-Gibbs sampler is more efficient than a standard Metropolis algorithm. The latter method samples all parameters simultaneously, while the former samples parameters individually. Under the Metropolis algorithm, if even one parameter suggestion is highly improbable, then  the entire vector of parameters is likely to be rejected. Therefore, a high-dimensional Markov chain may take an extremely long time to walk through parameter space and converge to the posterior distribution. By contrast, the hybrid-Gibbs sampler is much more efficient in our high-dimensional Markov chain (there are 2 model parameters, and 44 tangential velocity parameters for the Milky Way data discussed below). The parameters $M_{\text{tot}}$ and $a$ are sampled simultaneously, based on the current $v_t$ parameters, and the $v_t$ parameters are sampled individually based on the current values of all the other parameters.

Using the hybrid-Gibbs sampler for the $v_t$'s allows us to obtain a probability distribution for each $v_t$ parameter efficiently. We can look at the probability distribution for each $v_t$ and make a prediction of the most probable $v_t$ value for each satellite. Although we find that the resulting $v_t$ distributions are quite diffuse, and a meaningful prediction of $v_t$ cannot be made from them, the hybrid-Gibbs sampler method is nevertheless efficient and in general does not affect the mass estimate of the Galaxy (as will be shown below). If we assume that the satellite is bound to the galaxy, then by setting eq.~\ref{eq:scriptE} to zero we obtain an upper limit on the tangential velocity,
\begin{equation}\label{eq:vmax}
v_{t,max} = \sqrt{ 2\Psi(r) - v_r^2}.
\end{equation}

\subsection{Prior Probabilities}\label{sec:priors}
In a Bayesian analysis, the choice of a prior can be thought of as a chance for the researcher to state plainly and explicitly the prior assumptions. When little is known about the problem at hand, it is common to use a \emph{noninformative} prior, so that the information contained in the likelihood is not overwhelmed by information contained in the prior.

In this preliminary analysis, we use uniform priors for all of the model parameters because we assume little about the mass and scale of the system. The uniform prior for each parameter $\theta$ is,
\begin{equation}
	p(\theta) = \frac{1}{\theta_{max} - \theta_{min}}
\label{eq:unif-prior}
\end{equation}
where $\theta_{min}$ and $\theta_{max}$ are the lower and upper bounds of the uniform distribution. In practice, the parameters are sampled in the natural log-space to ensure that the total mass and scale radius are always positive. The Markov chain values are then exponentiated before examination of the posterior. The bounds $(\theta_{min}, \theta_{max})$ that we use for $M_{tot}$ and $a$ are on the order of $(10^8, 10^{15})\msun$ and $(10^{-2}, 10^6)$kpc respectively.

When the tangential velocities are treated as nuisance parameters and sampled in the Markov chain, they too require a prior. The tangential velocity is a 2-dimensional vector on the plane of the sky, so we would like the prior on $v_t$ to be uniform in $v^{2}_{t}$. Because we are sampling $v_t$ and not the squared tangential velocity, the uniform prior on $v^{2}_{t}$ needs to be transformed to one for $v_t$. 

Here we use Jeffreys' invariance principle \citep{jeffreys1939}. Suppose a parameter $\theta$ has a prior distribution $p(\theta)$, and that a one-to-one transformation is subsequently performed on $\theta$ such that $\phi=h(\theta)$. Then the Jeffrey's  prior for $\phi$ which expresses the same belief as that of $p(\theta)$ is
\begin{equation}
	p(\phi)=p(\theta)\left|\frac{d\theta}{d\phi}\right|=p(\theta)\left|h'(\theta)\right|^{-1}
\label{eq:invarianceprinc}
\end{equation}
\citep{gelman2003}.

Following equation~\ref{eq:invarianceprinc}, let $\theta=v^{2}_t$ and $\phi=v_t$, so that $h(\theta) = \sqrt{v^{2}_t}$. Then the prior on $v_t$ is given by
\begin{equation}
  p(v_t)=\frac{2v_t}{v^2_{t,max} - v^2_{t,min}}.
\label{eq:p(vt)}
\end{equation}
The minimum tangential velocity, $v_{t,min}$, is zero, and the maximum tangential velocity, $v_{t.max}$,  is a large constant. Note that if a value of $v_t$ that makes a particle unbounded is suggested in the Markov chain, it will be rejected via the likelihood.

\subsection{MCMC chains and Assessing convergence}\label{sec:convergence}

The computer code created for this research is written for the \R Project for Statistical Computing (\textbf{R}), an open source software environment for statistical computing and graphics \citep{R} with many well-developed and efficient statistical diagnostic tools. Recently, \R has gained popularity in astronomy and the field of astrostatistics \citep[e.g.][]{feigelson2012}, and our code is being developed into an \R package, called Galactic Mass Estimator (hereafter GME).

GME takes data of the form $(r, v_r, v_t)$ in Galactocentric coordinates, allows the user to select one of five DFs, constructs three Markov chains in parallel, and then combines them into a final, single chain after convergence conditions have been met. The result is a single Markov chain that represents samples from the posterior distribution for the model parameters, given the data. The SNOW package (\cite{snow}) is used for parallel computing, and the CODA package \citep{coda} is used for convergence diagnostics.

Many diagnostics to assess convergence of a Markov chain have been developed, and most of these methods use multiple chains. One advantage of running multiple chains is that the initial values can be dispersed widely in the parameter space, and then convergence can be approximated when they appear to reach a common stationary distribution. This approach allows more exploration of the parameter space and makes it less likely for a local maximum to be mistaken for the mode of the posterior. Furthermore, using multiple chains on the same data set allows estimates of convergence to be obtained in a more reliable and quantitative manner than a single chain. \cite{gelman1992} suggest using the statistic $\widehat{R}$ to assess the mutual convergence of parallel chains:
\begin{equation}
	\widehat{R}=\sqrt{\frac{\widehat{\text{var}}^{+}\left(\psi|y\right)}{W}}.
	\label{eq:Rhat}
\end{equation}
In equation~\ref{eq:Rhat}, $\widehat{\text{var}}^{+}\left(\psi|y\right)$ is the \emph{marginal posterior variance of the estimand [parameter]}, which is essentially a weighted average of the within-chain variance $W$, and between-chain variance $B$ \citep[see][for more details]{gelman2003}. In practice, $\widehat{R}$ is calculated for each parameter separately; to assume convergence, \cite{gelman2003} recommend a value of $\widehat{R}<1.1$ for every parameter, and this is the criterion we use. The $\widehat{R}$ statistic is available in the \R Software Statistical Computing Language in the CODA package \citep{coda}.

In this work, the three Markov chains' starting values are widely dispersed in the parameter space, and each chain is run for $i=10^3$ iterations. After this initial run, $\widehat{R}$ is calculated for each parameter. If any of the parameters have $\widehat{R}>1.1$, then it is assumed that the chains have not converged, and the last parameter values in each chain are used as the initial parameters in three new chains. The process is repeated until all parameters across all three chains have $\widehat{R}<1.1$, at which point convergence is assumed. The final sample of the posterior distribution is created by combining the last halves of the three chains, thus providing 1500 parameter vector samples. Prior to this final step, however, we check the effective sample sizes of the Markov chains.

In general, the draws in a Markov chain are not truly independent; some autocorrelation exists in the sequence of samples \citep{gelman2003}. The effective sample size is the equivalent number of independent samples:
\begin{equation}
 n_{\text{eff}} = mn \frac{\widehat{\text{var}}^{+}\left(\psi|y\right)}{B}
 \label{eq:neff}
\end{equation}
where $m$ is the number of parallel Markov chains and $n$ is the number of draws in each chain \citep{gelman2003}. If the draws in all chains were perfectly independent, then the number of independent draws would be $mn$. However, the draws within a chain are general autocorrelated, and so $B>\widehat{\text{var}}^{+}\left(\psi|y\right)$, and $n_{eff}<mn$. An effective sample size of at least 100 is necessary to obtain reliable first-moment statistics such as the mean and median, while an effective sample size over 200 is needed for second-order moments.

In our code, we use the \emph{effectiveSize} function in the CODA package \citep{coda}, and find that all parameters had effective sample sizes greater than 300 for all models when $i=10^4$. An acceptance rate between 20 and 30\% is required for the $v_t$ parameters, and an acceptance rate of 30-40\% is required for the model parameters. The final chain (15000 samples) for each model is visually inspected to ensure that the three chains did not reach very different maxima in the posterior distribution.

\section{Simulations}\label{sec:sims}

Prior to data analysis, we explored the statistical properties of our Bayesian estimates under repeated sampling, using simulation. Unfortunately, a Bayesian analysis cannot be repeated when the data is real (i.e. for a single data set). A Bayesian analysis can be repeated, however, with simulated data sets produced from the same DF, and analysed independently using the same model. By examining the range of parameter estimates, the average behaviour of the model on this type of data can be explored. Simulations and analyses of trivial cases (i.e. when the model and data have the same distribution) are also an effective way to test code.

It is expected that when the simulated data come from the same DF as that of the model, then on average the Bayesian parameter estimates should be correct. Furthermore, quantities like uncertainties and credible intervals should be reliable (e.g. a 50\% credible region should contain the true parameter values 50\% of the time). In contrast, when the simulated data comes from a different DF than that of the model, biases in the estimates may occur, and the credible regions may become unreliable (e.g. overconfident). Moreover, we need to investigate whether or not treating missing $v_t$ measurements as parameters affects other parameter estimates, regardless of whether or not the data and model share the same DF.

We simulate mock observations of 100 satellites orbiting a galaxy whose gravitational potential follows the Hernquist model.  The mock tracer observations include their distances $r$ and velocity components $v_r$ and $v_t$. We explore the effects of assuming an isotropic Hernquist model in the following three scenarios:
\begin{enumerate}\label{list:rt}
 \item isotropic data with complete velocity vectors\label{scen1}
 
 \item isotropic data with 50 unknown $v_t$ values\label{scen2}
 
 \item constant anisotropic data from the $\beta=0.5$ Henquist model, with 50 unknown $v_t$'s\label{scen3}

\end{enumerate}
For each scenario, we create 500 data sets with 100 particles each. The Bayesian analysis is performed on each set of data, as described in Sections~\ref{sec:Bayes}~-~\ref{sec:convergence}, yielding 500 Markov chains for each scenario. From these chains, statistics such as the mean parameter estimate and credible intervals are calculated.

For all the simulations, we use $M_{\text{tot}}=10^{12}$\msun ~and $a=15\kpc$~ to generate the data. Our choice for the total mass is based on many other studies that have shown the Milky Way's mass to be close to this order of magnitude. For numerical simplicity, the following units are used in our code: the gravitational constant $G \equiv 1$, $r$ is measured in kiloparsecs (kpc), velocity components are measured in $100$\kms, and mass is measured in $2.325\times10^{9}$\msun.

\section{Kinematic Data for the Milky Way}\label{sec:data}

\begin{table*}[t]
\begin{center}
\caption{Milky Way Kinematic Data}\label{tab:data}
\begin{tabular}{lrrrrrr|lrrrrrr}
\tableline\tableline
Object & $r$ & $v_{r}$ & $\Delta v_r$ & $v_{t}$ & $\Delta v_t$ & $\cos{\gamma}$ &
Object & $r$ & $v_{r}$ & $\Delta v_r$ & $v_{t}$ & $\Delta v_t$ & $\cos{\gamma}$ \\
& (kpc) & (\kms) &  & (\kms) &  &  & & (kpc) & (\kms) &  & (\kms) & \\
\tableline
\hline
 NGC 104 & 7 & 17.0 & 0.2 & 171.0 & 22.0 & 0.15 & NGC 6540 & 3 & 0.0 & 1.4 & -- & -- & -0.97 \\ 
  NGC 288 & 11 & 16.0 & 0.4 & 59.0 & 18.0 & 0.75 & NGC 6569 & 3 & -0.2 & 5.6 & -- & -- & 0.95 \\ 
  NGC 362 & 9 & 55.0 & 0.5 & 85.0 & 31.2 & 0.61 & NGC 6864 & 15 & -1.1 & 3.6 & -- & -- & 0.96 \\ 
  NGC 1851 & 16 & 186.0 & 0.6 & 170.0 & 42.4 & 0.89 & IC 1257 & 18 & -66.5 & 2.1 & -- & -- & 0.99 \\ 
  NGC 1904 & 18 & 93.0 & 0.5 & 83.0 & 44.7 & 0.94 & Arp 2 & 21 & 153.0 & 10.0 & -- & -- & 0.99 \\ 
  NGC 2298 & 14 & -58.0 & 1.3 & 100.0 & 52.7 & 0.89 & NGC 7492 & 25 & -97.4 & 0.6 & -- & -- & 0.95 \\ 
  Pal 3 & 85 & -247.0 & 8.4 & 242.0 & 121.5 & 1.00 & NGC 5824 & 26 & -117.7 & 1.5 & -- & -- & 0.98 \\ 
  NGC 4147 & 19 & 57.0 & 1.0 & 161.0 & 65.7 & 0.93 & Pal 13 & 27 & 192.4 & 0.3 & -- & -- & 0.95 \\ 
  NGC 4590 & 9 & -99.0 & 0.6 & 300.0 & 35.6 & 0.69 & NGC 5694 & 29 & -228.1 & 0.8 & -- & -- & 0.98 \\ 
  NGC 5024 & 18 & -106.0 & 4.1 & 250.0 & 86.5 & 0.90 & NGC 6229 & 30 & 22.6 & 7.6 & -- & -- & 0.96 \\ 
  NGC 5139 & 6 & -31.0 & 0.7 & 65.0 & 14.1 & 0.05 & Whiting 1 & 35 & -103.5 & 1.8 & -- & -- & 0.98 \\ 
  NGC 5272 & 12 & 2.0 & 0.4 & 164.0 & 24.5 & 0.75 & Pal 2 & 35 & -104.4 & 57.0 & -- & -- & 1.00 \\ 
  NGC 5466 & 16 & 254.0 & 0.3 & 216.0 & 66.8 & 0.88 & Pal 15 & 38 & 147.8 & 1.1 & -- & -- & 0.99 \\ 
  Pal 5 & 16 & -11.0 & 16.0 & 62.0 & 38.0 & 0.95 & NGC 7006 & 39 & -185.2 & 0.4 & -- & -- & 0.98 \\ 
  NGC 5897 & 7 & 49.0 & 1.0 & 138.0 & 59.4 & 0.79 & Pyxis & 41 & -195.2 & 1.9 & -- & -- & 0.98 \\ 
  NGC 5904 & 6 & -313.0 & 0.5 & 234.0 & 39.6 & 0.33 & Pal 14 & 72 & 165.4 & 0.2 & -- & -- & 1.00 \\ 
  NGC 6093 & 3 & 60.0 & 4.1 & 85.0 & 28.2 & 0.67 & NGC 2419 & 90 & -26.4 & 0.5 & -- & -- & 1.00 \\ 
  NGC 6121 & 6 & -58.0 & 0.4 & 25.0 & 22.6 & -0.91 & Eridanus & 95 & -141.0 & 2.1 & -- & -- & 1.00 \\ 
  NGC 6144 & 3 & 109.0 & 1.1 & 137.0 & 33.3 & 0.47 & Pal 4 & 111 & 50.5 & 2.1 & -- & -- & 1.00 \\ 
  NGC 6171 & 4 & 20.0 & 0.3 & 156.0 & 36.9 & -0.29 & AM 1 & 125 & -41.6 & 20.0 & -- & -- & 1.00 \\ 
  NGC 6205 & 8 & 279.0 & 0.9 & 129.0 & 35.0 & 0.48 & Fornax & 140 & -31.8 & 1.7 & 196.0 & 29.0 & 1.00 \\ 
  NGC 6218 & 5 & -21.0 & 0.6 & 168.0 & 22.0 & -0.46 & LeoI & 261 & 167.9 & 2.8 & 101.0 & 34.4 & 1.00 \\ 
  NGC 6254 & 5 & -53.0 & 1.1 & 178.0 & 28.3 & -0.59 & LMC & 49 & 93.2 & 3.7 & 346.0 & 8.5 & 0.99 \\ 
  NGC 6341 & 9 & 70.0 & 1.7 & 46.0 & 26.2 & 0.61 & SMC & 60 & 6.8 & 2.4 & 259.0 & 17.0 & 0.99 \\ 
  NGC 6362 & 5 & -40.0 & 0.6 & 134.0 & 20.5 & 0.25 & Sculptor & 87 & 79.0 & 6.0 & 198.0 & 50.0 & 1.00 \\ 
  NGC 6397 & 6 & 18.0 & 0.1 & 166.0 & 16.3 & -0.83 & Draco & 92 & -98.5 & 2.6 & 210.0 & 25.0 & 1.00 \\ 
  NGC 6584 & 6 & 150.0 & 15.0 & 185.0 & 55.9 & 0.88 & BootesI & 57 & 106.6 & 1.0 & -- & -- & 0.99 \\ 
  NGC 6626 & 3 & 8.0 & 1.0 & 172.0 & 26.4 & -0.87 & BootesII & 43 & -115.6 & 5.0 & -- & -- & 0.98 \\ 
  NGC 6656 & 5 & 172.0 & 0.6 & 214.0 & 31.9 & -0.94 & CanesVenaticiI & 219 & 76.8 & 1.0 & -- & -- & 1.00 \\ 
  NGC 6712 & 4 & 208.0 & 0.6 & 132.0 & 21.5 & -0.08 & CanesVenaticiII & 150 & -96.1 & 1.0 & -- & -- & 1.00 \\ 
  NGC 6752 & 5 & -19.0 & 1.5 & 200.0 & 11.4 & -0.50 & Carina & 102 & 14.3 & 1.0 & -- & -- & 1.00 \\ 
  NGC 6779 & 9 & 172.0 & 0.9 & 39.0 & 58.1 & 0.63 & ComaBernices & 45 & 82.6 & 5.0 & -- & -- & 0.98 \\ 
  NGC 6809 & 4 & -181.0 & 0.4 & 119.0 & 30.4 & -0.49 & Hercules & 141 & 142.9 & 1.0 & -- & -- & 1.00 \\ 
  NGC 6838 & 7 & 3.0 & 0.2 & 180.0 & 17.8 & -0.05 & LeoII & 235 & 26.5 & 8.0 & -- & -- & 1.00 \\ 
  NGC 6934 & 12 & -305.0 & 1.6 & 124.0 & 93.0 & 0.86 & LeoIV & 154 & 13.9 & 1.0 & -- & -- & 1.00 \\ 
  NGC 7078 & 10 & -74.0 & 0.6 & 141.0 & 34.7 & 0.70 & LeoV & 175 & 62.3 & 3.0 & -- & -- & 1.00 \\ 
  NGC 7089 & 10 & 46.0 & 0.9 & 331.0 & 63.9 & 0.74 & Sagittarius & 16 & 166.3 & 60.0 & -- & -- & 0.93 \\ 
  NGC 7099 & 7 & 14.0 & 1.0 & 120.0 & 30.8 & 0.46 & Segue1 & 28 & 113.5 & 1.0 & -- & -- & 0.97 \\ 
  NGC 5634 & 21 & -0.8 & 6.6 & -- & -- & 0.95 & Segue2 & 41 & 39.7 & 1.0 & -- & -- & 0.99 \\ 
  NGC 6284 & 8 & 0.3 & 1.7 & -- & -- & 0.98 & Sextans & 89 & 78.2 & 1.0 & -- & -- & 1.00 \\ 
  NGC 6356 & 7 & 0.6 & 4.3 & -- & -- & 0.97 & UrsaMajorI & 101 & -8.8 & 1.0 & -- & -- & 1.00 \\ 
  NGC 6426 & 14 & -0.5 & 23.0 & -- & -- & 0.96 & UrsaMajorII & 36 & -36.5 & 2.0 & -- & -- & 0.99 \\ 
  NGC 6441 & 4 & -0.0 & 1.0 & -- & -- & 0.95 & UrsaMinor & 77 & -89.8 & 8.0 & -- & -- & 1.00 \\ 
  NGC 6453 & 4 & -0.9 & 8.3 & -- & -- & 0.98 & Willman1 & 42 & 33.7 & 2.0 & -- & -- & 0.98 \\ 
  
\tableline
\end{tabular}

\tablecomments{Columns from left to right: objects' names, Galactocentric distance, radial velocity, uncertainty in radial velocity, tangential velocity, uncertainty in tangential velocity,and $\cos{\gamma}$. All data are in Galactocentric coordinates ($r$, $v_r$, $v_t$) as described in Section~\ref{sec:newmethods}, with the exception of GCs and DGs that lack tangential velocities (see text). Conversions from line-of-sight and proper motion measurements to Galactocentric measurements were completed by the studies mentioned in Section~\ref{sec:data}.}
\end{center}
\end{table*}

In principle, any well-defined object orbiting the Galaxy with a measured distance from the Galactic center and at least one velocity measurement may be used to estimate the mass of the Milky Way. In this work, as a first run, we use only globular clusters (GCs) and dwarf galaxies (DGs). It is possible to measure the proper motions and line-of-sight velocities of these tracers in the Galaxy's halo, and to convert these measurements into Galactocentric coordinates. Indeed, many of the kinematic measurements and conversions have already been made \citep[e.g.][]{1999Dinescu, 2010Dinescu, 2013Dinescu, boylan2013}. However, the proper motions of many GCs and DGs have yet to be measured, and so the conversion from our frame of reference to a Galactocentric one cannot be performed. Nevertheless, the line-of-sight velocities of these objects are available, and could contain useful information about the Galaxy's mass profile. Thus, we incorporate some of this incomplete data into our analysis.

The data used in this research are in Galactocentric coordinates (see Table~\ref{tab:data}). The first 59 objects are GCs, and the last 29 are DGs. Note that 26 GCs and 18 DGs listed do not have tangential velocities, because they have no proper motion measurements. The Galactocentric radial velocities for these data must be approximated, for which we assume $v_r\approx v_{\text{los}}$. We use this approximation only for objects with $\lvert \cos{\gamma} \rvert \geq 0.95$ (where $\gamma$ is the angle subtended by the line connecting the Sun and the Galactic Centre, from the object), guaranteeing that any further adjustment to $v_{\text{los}}$ will be small. We also exclude the following clusters, even though they have $\lvert\cos{\gamma}\rvert \geq 0.95$, because they are either associated with the Sagittarius dwarf galaxy or their measurements suffer from high extinction: Djorg 1, NGC 6401, NGC 6715, NGC 6544, NGC 6715, Pal 6, Terzan 1, Terzan 6, Terzan 7, and Terzan 8.

The GC data are taken from \cite{1999Dinescu}, \cite{2010Dinescu}, \cite{2013Dinescu}, and \cite{1996harrisPaper}, while data for six of the dwarf galaxies are taken from \cite{sohn2013} (Leo I), \cite{pryor2014} (Draco), and \cite{boylan2013} (Fornax, LMC, SMC, and Sculptor). The rest of the dwarf galaxy data, which do not have tangential velocities, are from the compilation given in \cite{watkins2010} and references therein. Uncertainties in the \cite{watkins2010} dwarfs' $v_r$ values are taken from the HyperLeda Catalogue \citep{paturel2003}, with the exception of those for Coma Berenices, Sagittarius, and Sextans, which are taken from \cite{SimonGeha2007}, \cite{ibata1997}, and \cite{walker2006} respectively. The $r$-values in Table~\ref{tab:data} are based on mean magnitudes of RR Lyrae and horizontal branch stars, and are uncertain to typically $5\%$ \citep[see][]{1996harrisPaper}. The uncertainties associated with $r$ and $v_r$, and the differences in the LSR assumed motion used among the different studies are $\lessapprox15$~km/s, and thus unimportant compared to the uncertainties associated with the $v_t$ values.

In the following analysis, we specifically assume (a) a spherical Hernquist-like or Jaffe-like halo potential, (b)
equal weights for all data points, (c) no net rotation of the halo, and (d) that all tracers are bound to the Galaxy.

\section{Results}\label{sec:results}

\subsection{Simulation Results}

\begin{figure*}[!t]
\centering
	\includegraphics[trim=0cm 0.5cm 0.5cm 2cm, totalheight=0.4\textheight]{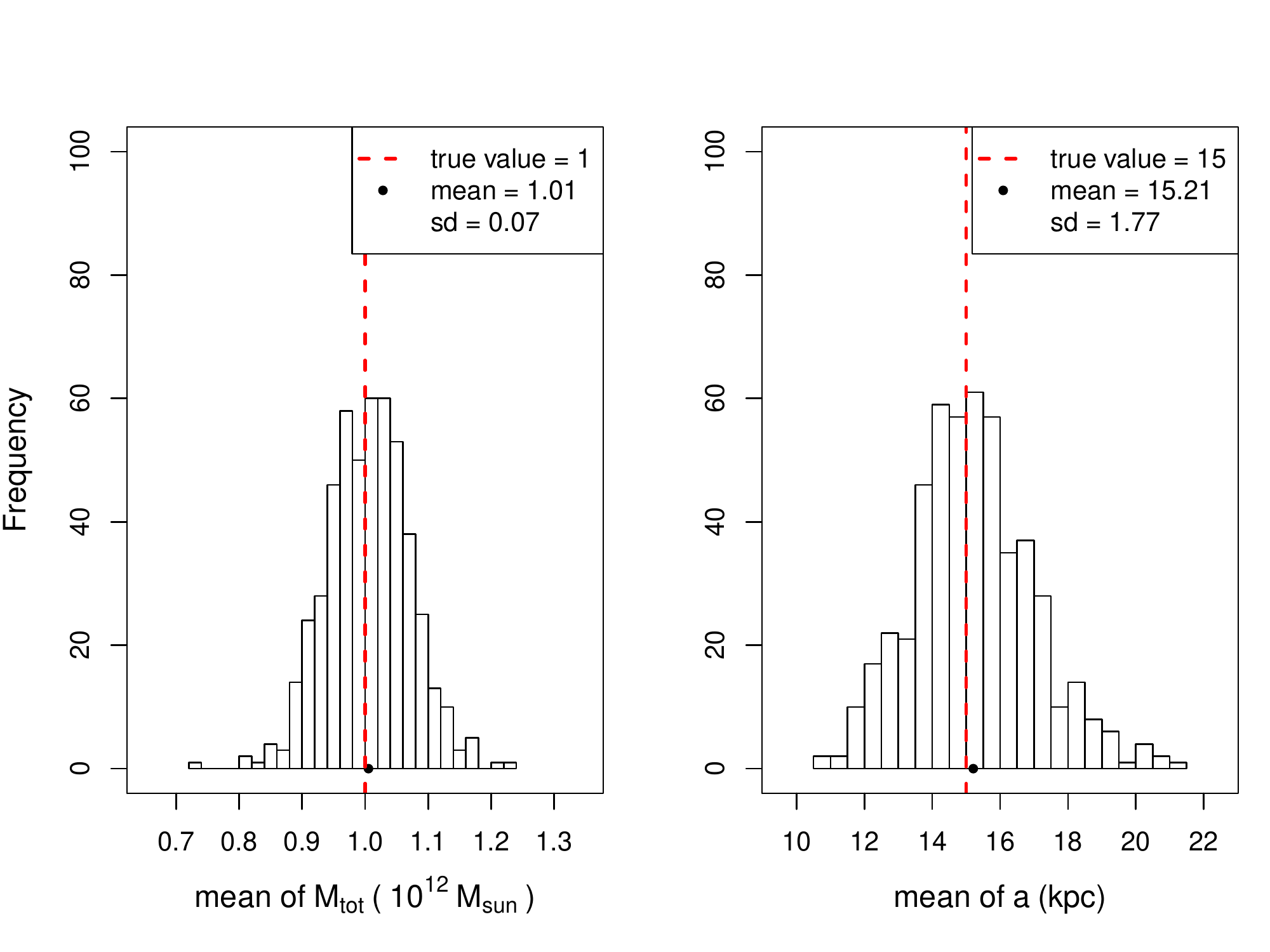}
	\caption{Empirical distribution of $M_{tot}$ and $a$ estimates from simulated data analysis. Black points and red dashed lines show the mean of the estimates and the true value of the parameter respectively. The standard deviation of the estimates is 0.07 and 1.77 for $M_{tot}$ and $a$ respectively.}
	\label{fig:histiso}
\end{figure*}

Figure~\ref{fig:histiso} shows the distribution of the mean parameter estimates from scenario~\ref{scen1}. Black dots are the mean of the estimates, and red dashed lines are the true parameter values. On average, the estimates are unbiased within one standard deviation (sd), and the sd of the chains is roughly equal to the sd of the estimates.

\begin{figure}
\centering
	\includegraphics[trim=0cm 0cm 1cm 0cm, clip=true, totalheight=0.4\textheight]{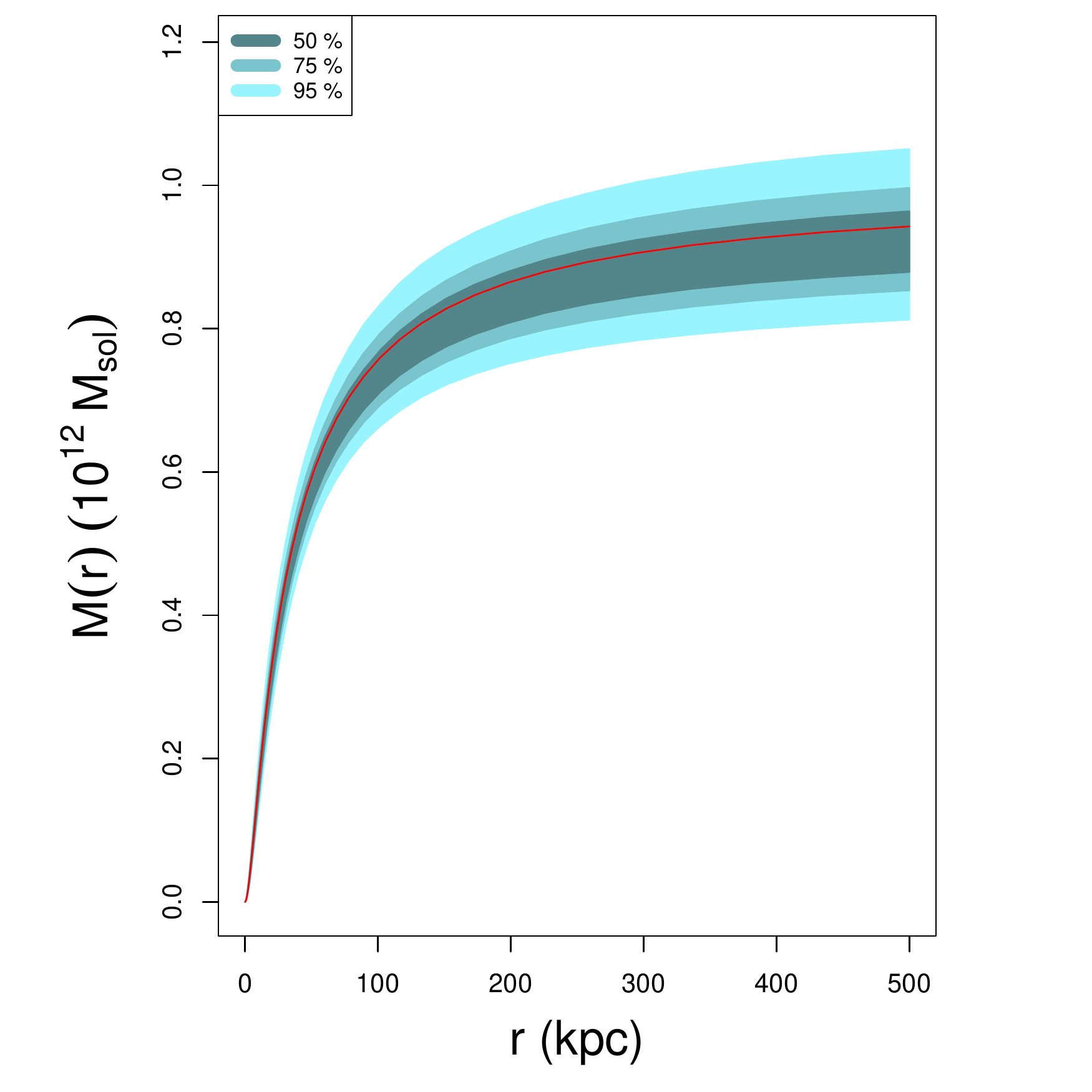}
	\caption{Example of predicted and true mass profile from analysis of simulated data. The true $M(r)$ profile is shown in red, and the 50, 75, and 96\% credible regions are shown as shades of teal. Note: this is the result from one analysis (i.e. one data set). }
	\label{fig:Mr}
\end{figure}

Because the Markov chain represents the posterior distribution, we can also calculate credible regions --- Bayesian analogues of confidence intervals --- for the $M(r)$ profile. An example of the mass profile credible regions for one data set is shown in Figure~\ref{fig:Mr}, where shades of teal show the 50, 75, and 95\% credible intervals as a function of $r$. Credible regions are found by calculating $M(r)$ at several different $r$ values, for every set of parameters in the Markov chain. The true $M(r)$ profile is the solid red line, calculated from eq.~\ref{eq:M(r)} and the true $M_{tot}$ and $a$.  We find that the credible regions are reliable when the DF of the assumed model and the DF of the data are the same. For example, the true $M(r)$ curve fell within the 75\% credible region seventy-five percent of the time over the course of the 500 analyses for scenario~\ref{scen1}.

\begin{figure*}[h!]
\centering
\includegraphics[trim=0cm 0cm 0cm 1.5cm, clip=true, totalheight=0.45\textheight]{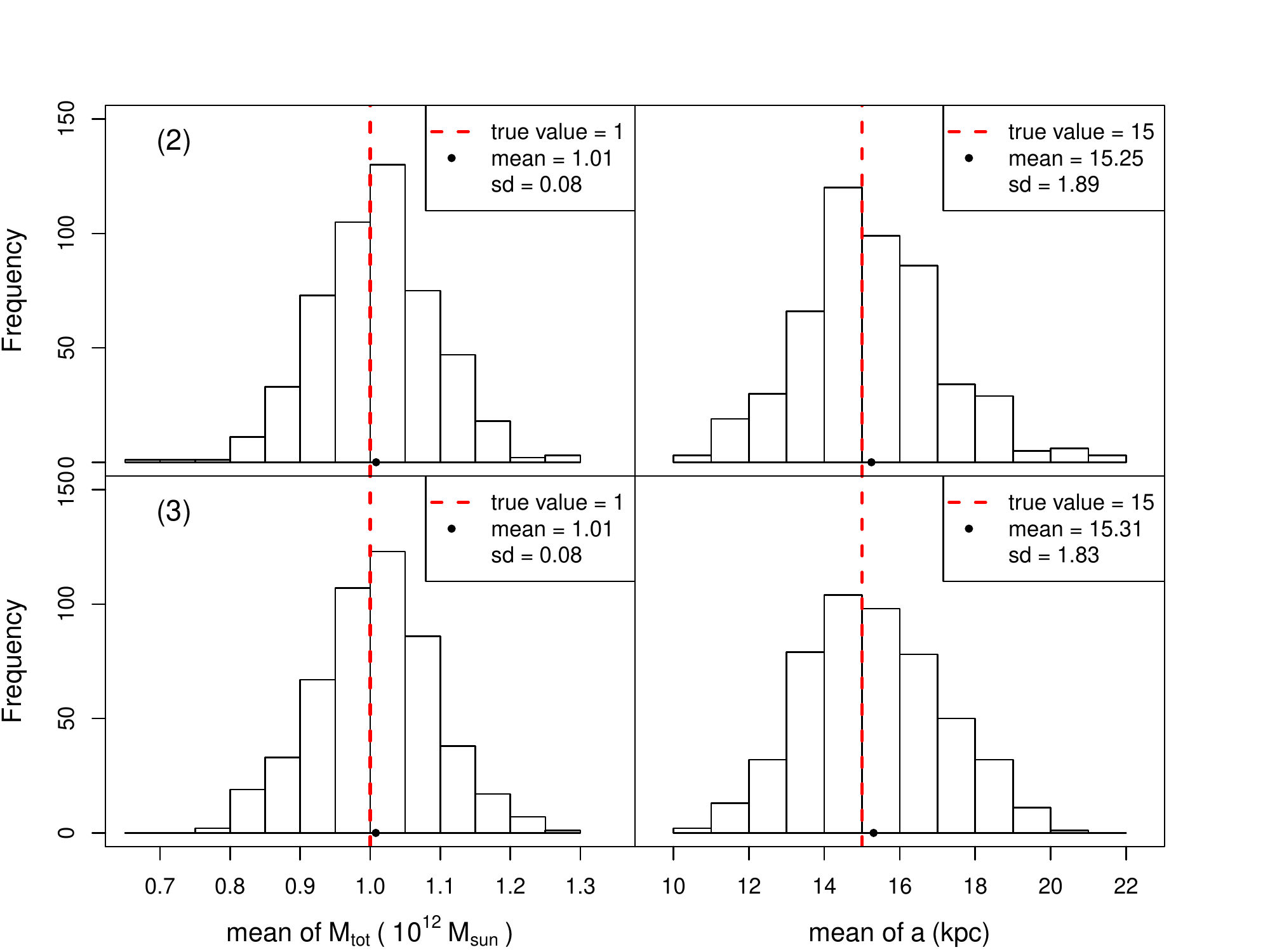} 
	\caption{Empirical distribution of $M_{tot}$ and $a$ estimates from simulated data analysis, with 50 tangential velocities removed. The top panels are for scenario~(\ref{scen2}) and the bottom, (\ref{scen3}). The black points and red dashed lines show the mean of the estimates and the true value of the parameters respectively. The standard deviations of the estimates in scenario~(\ref{scen2}) are  $0.08\times10^{12}$\msun~ and $1.9$\kpc, while the standard deviations are $0.08\times10^{12}$\msun~ and $1.8$\kpc~ in (\ref{scen3}).}
	\label{fig:histisogibbs}
\end{figure*}

\begin{figure*}[h!]
\centering
\parbox{3in}{\includegraphics[trim=1cm 0cm 0cm 0cm, clip=true, totalheight=0.4\textheight]{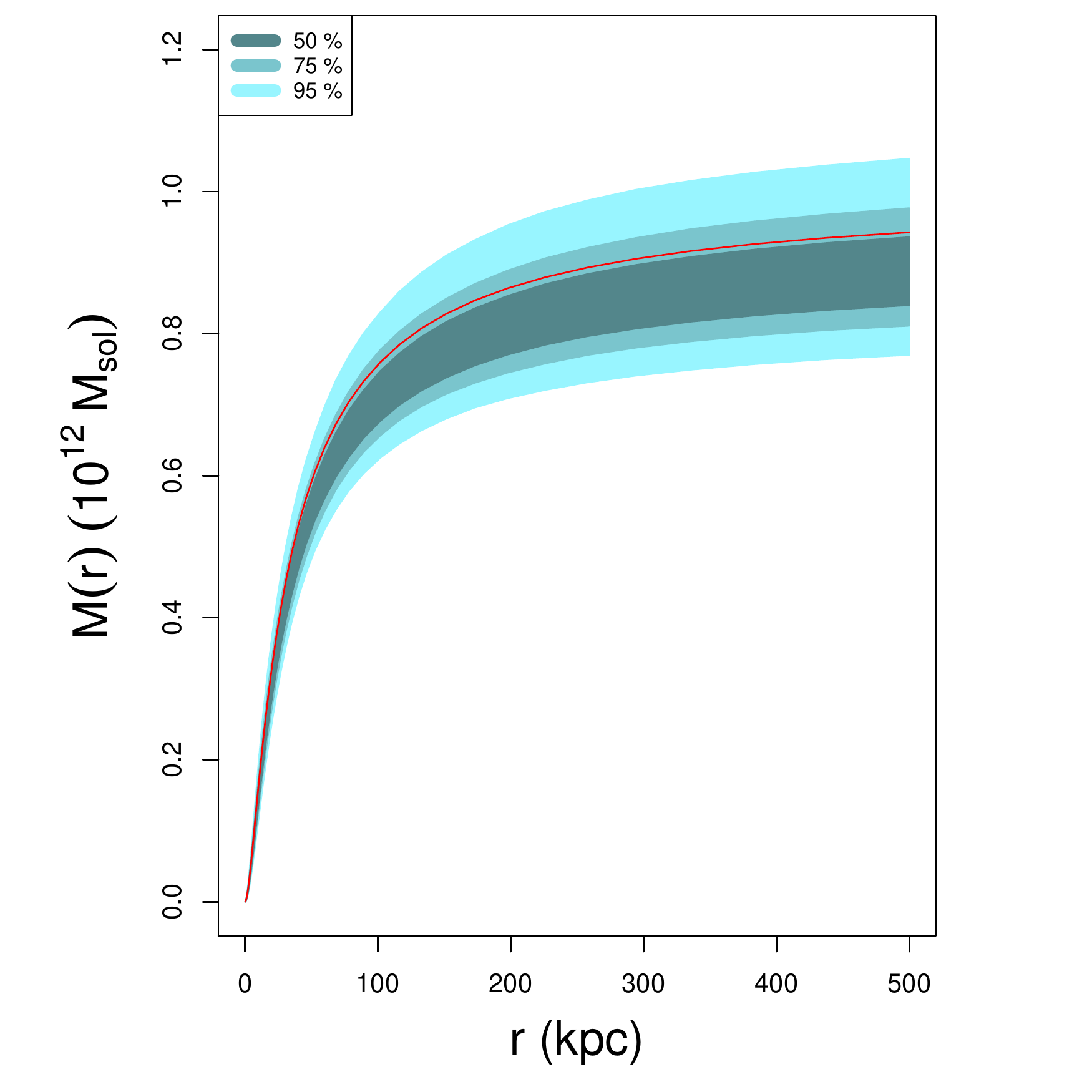}
}%
\qquad
\begin{minipage}{3in}%
\includegraphics[trim=1cm 0cm 0cm 0cm, clip=true, totalheight=0.4\textheight]{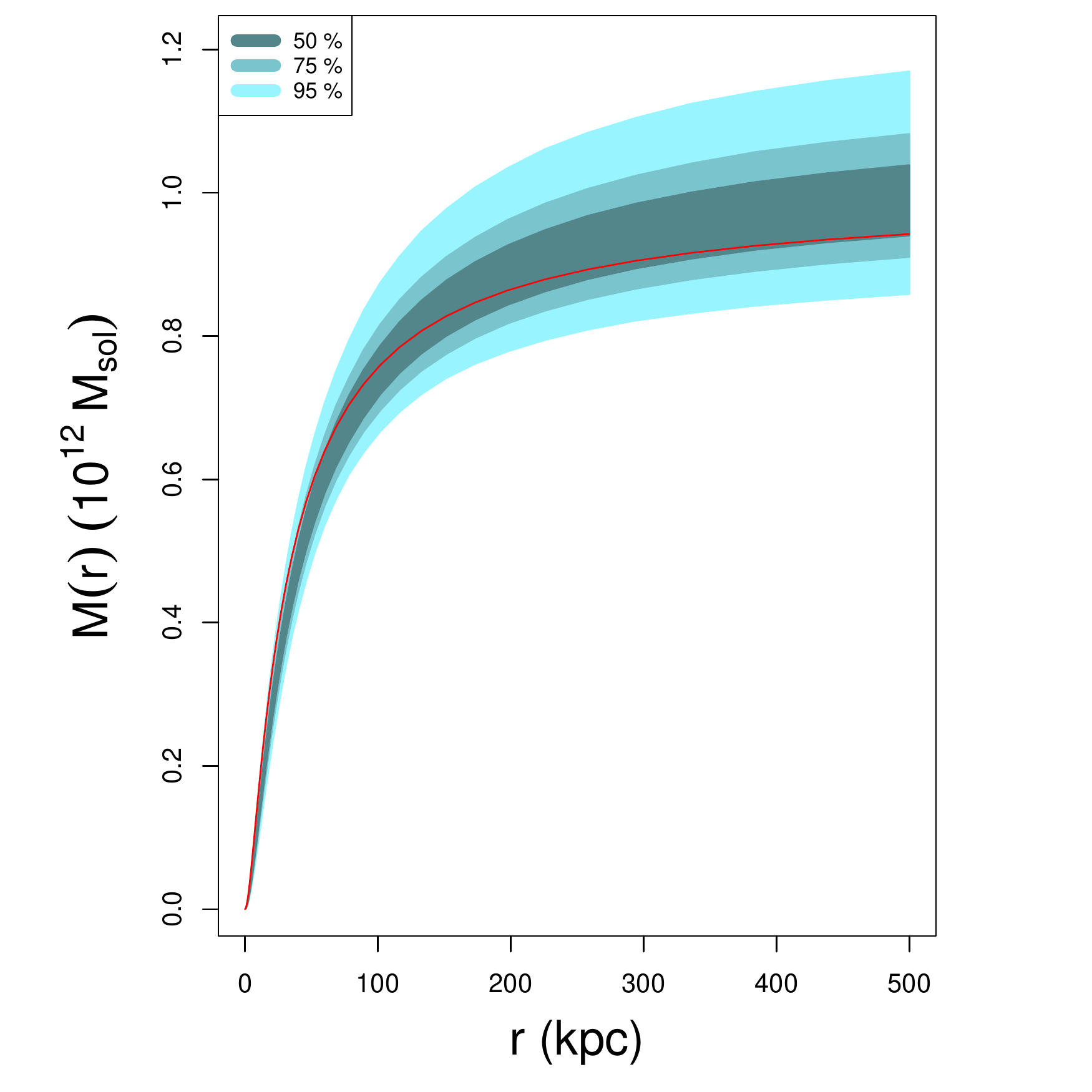}

\end{minipage}
\caption[This is where the title goes for the TOC.]{Example cumulative mass profile when 50 $v_t$'s are unknown, and an isotropic Hernquist model is assumed. The true profile is shown as a solid red line, and the credible regions are shown as shades of teal. The left profile is scenario (\ref{scen2}) and the right is scenario (\ref{scen3}).}
	\label{fig:Mrisogibbs}
\end{figure*}

In scenario~\ref{scen2}, we randomly remove 50 $v_t$'s from each data set, and treat them as parameters in the analysis. We find a very small positive bias in both the $M_{tot}$ and $a$ estimates, as shown in the top two panels in Fig.~\ref{fig:histisogibbs}. The bias is insignificant, as the means of the estimates ($1.01\times10^{12}$\msun~ and $15.2$\kpc) are within one standard deviation of the distribution ($0.08\times10^{12}$\msun~ and $1.9$\kpc~ respectively). The slight although insignificant positive bias suggests that the median may be a better estimate of the mass than the mean, but the median is almost identical to the mean in all cases. 

In scenario~\ref{scen3} recall that an isotropic model is assumed, but the data sets in scenario~\ref{scen3} have constant anisotropic velocity dispersions $\beta=0.5$. Despite the data and model having different DFs, the estimates show only a slight positive bias; the true $M_{tot}$ and $a$ are still within one standard deviation of the distribution (see Fig.~\ref{fig:histisogibbs}). The mean of the estimates for $M_{tot}$ and $a$ are $1.01\pm0.08\times10^{12}$\msun and $15.3\pm1.8$\kpc~respectively.

Examples of mass profile credible regions from scenarios~\ref{scen2} and ~\ref{scen3} are presented in Fig.~\ref{fig:Mrisogibbs}. Note that the introduction of $v_t$ parameters tends to increase the width of the credible regions at all $r$ values compared to Fig.~\ref{fig:Mr}. In scenario~\ref{scen2}, we find the credible regions to be slightly over confident for values of $17<r<35$kpc, with the true $M(r)$ falling within the 50, 75, and 95\% regions 48, 73, and 93\% of the time over the 500 analyses. At all other $r$ values, however, the credible regions are reliable. In scenario~\ref{scen2}, the credible regions are slightly lower than the true cumulative mass profile; the opposite is the case for scenario~\ref{scen3}, but for both the true curve still lies in the 75\% credible region for most $r$ (see Fig.~\ref{fig:Mrisogibbs}). We reiterate, however, that Fig.~\ref{fig:Mrisogibbs} are examples of $M(r)$ profiles from a single data set and analysis. Over  500 analyses we find that the 50, 75, and 95\% credible regions do contain the true $M(r)$ 50, 75, and 95\% of the time, in both scenarios~\ref{scen2} and \ref{scen3}, for almost all $r$.

\subsection{Milky Way Results}
Assuming an isotropic Hernquist model, and using all the kinematic data from Table~\ref{tab:data}, we find a mean \Mtot of $1.55\pm 0.08 \times 10^{12}$ \msun~and a scale radius of $16.9\pm2.3$~kpc, where the uncertainties are the standard deviations of the parameters in the Markov chain. The $95\%$ credible regions for $M_{tot}$ and $a$ are $(1.42, 1.73) \times 10^{12}$ \msun~ and $(12.8, 21.7)$~kpc respectively. We also report the mean \Mtot and scale radius, with uncertainties of one standard deviation, when the other models are assumed (Table~\ref{tab:results}). The mass estimates and scale radii vary only slightly between Hernquist models. The Jaffe model's mass is similar, but the Jaffe scale radius radius cannot be compared directly to that of Hernquist because they have physically different definitions.

\begin{figure}[h]
	\includegraphics[trim=0cm 0cm 0cm 0cm, clip=true, totalheight=0.4\textheight]{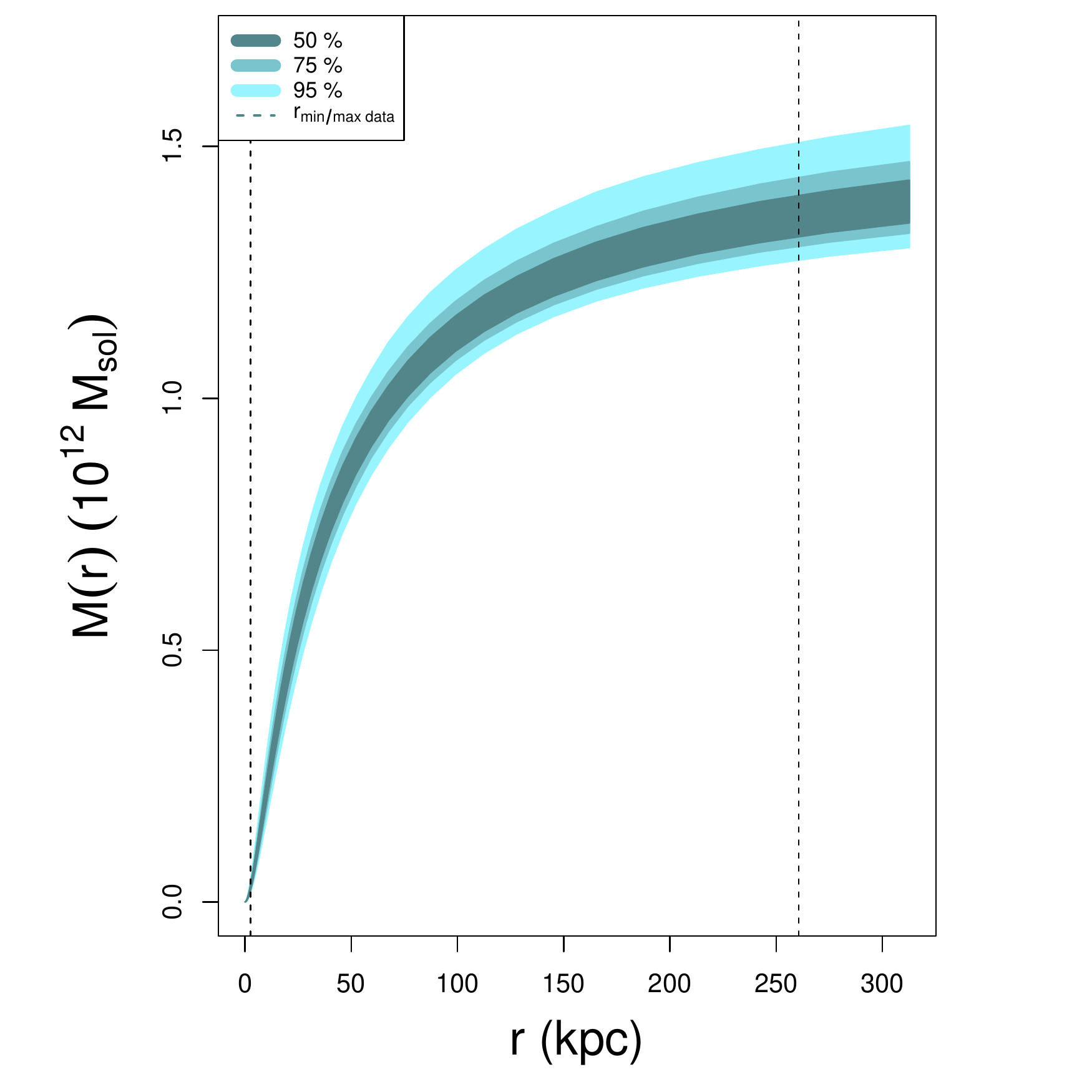}
	\caption[This is where the title goes for the TOC.]{Mass profile credible regions assuming a Hernquist model with isotropic velocity dispersion. The dashed lines indicate the location of NGC 6540 and Leo I (the closest and furthest objects from the Galactic center respectively in our dataset).}
	\label{fig:credreg}
\end{figure}

The mass profile credible regions are shown in Fig.~\ref{fig:credreg}. The innermost dark region corresponds to the 50\% credible region. 
The vertical dashed lines show the extent of the data, with NGC~6540 and Leo~I being the closest and furthest objects from the Galactic center respectively. The mass contained within the distance of Leo~I is $1.37^{+0.14}_{-0.10}\times10^{12}$\msun, where the uncertainties correspond to the 95\% credible interval.

\begin{table*}
\begin{center}
 \caption{\textbf{Parameter Estimates for the Milky Way.}}
 \label{tab:results}
 \begin{tabular}{lllllll}
 \tableline
 \T  
   & \multicolumn{2}{|c|}{\bf{All Data}} & \multicolumn{2}{|c|}{\bf{Without Pal 3}} & \multicolumn{2}{|c|}{\bf{Without Draco}}\\
 \tableline
 \T
 Model - $\sigma^2$ & \Mtot & Scale Radius& \Mtot & Scale Radius & \Mtot & Scale Radius \\
  & ($10^{12}$\msun) & (kpc) &  ($10^{12}$\msun) & (kpc) &  ($10^{12}$\msun) & (kpc)  \\
	\tableline
	\T
	
  H - iso & 1.55 $\pm$ 0.08 & 16.9 $\pm$ 2.3 & 1.36 $\pm$ 0.07 & 16.7 $\pm$ 2.2 & 1.55 $\pm$ 0.08 & 16.8 $\pm$ 2.3 \\ 
  H - OM & 1.52 $\pm$ 0.08 & 16.7 $\pm$ 2.2 & 1.34 $\pm$ 0.06 & 16.4 $\pm$ 2.1 & 1.52 $\pm$ 0.08 & 16.6 $\pm$ 2.2 \\ 
  H - $\beta=0.5$  & 1.47 $\pm$ 0.07 & 12.1 $\pm$ 1.9 & 1.31 $\pm$ 0.06 & 12.3 $\pm$ 1.9 & 1.46 $\pm$ 0.07 & 11.9 $\pm$ 1.8 \\ 
  Jaffe - iso & 1.61 $\pm$ 0.09 & 47.7 $\pm$ 8.5 & 1.38 $\pm$ 0.06 & 48.9 $\pm$ 8.7 & 1.57 $\pm$ 0.08 & 45.2 $\pm$ 8.2 \\

 \tableline
 \end{tabular}
 
 \tablecomments{In the first column, the first three models are of the Hernquist type, with isotropic, OM-type anisotropy, and constant anisotropy. The last row shows the results of assuming an isotropic Jaffe model. Uncertainties are one standard deviation of the posterior distribution.}
 \end{center}
\end{table*}

Some satellites may have a large effect on the mass estimate of the Galaxy. Leo I, for example, remained a contentious object for many years, because it is at a large distance from the Galactic center and it was unclear whether or not it is bound to the MW. Recently, however, \cite{boylan2013} showed that Leo I is likely bound to the MW. Furthermore, when Leo I's proper motion is taken into account,  the object has little effect on the mass estimate of the Galaxy \citep{wilkinsonevans1999}. We run our analysis assuming the isotropic Hernquist model both with and without Leo I, and we also find that it has no effect within error on $M_{tot}$. When Leo I is removed from the analysis, \Mtot$=1.52\pm 0.07 \times 10^{12}$\msun ~and $a=16.2\pm2.2$~kpc, very similar to the values obtained when Leo I is present. 

The five other dwarf galaxies in our data set that have measured proper motions are Draco, Fornax, Sculptor, and the Large and Small Magellanic Clouds (hereafter LMC and SMC). We obtain parameter estimates assuming an isotropic Hernquist model with each of these dwarfs removed, and find that the mass and scale radii do not change within error in any case.

Another object to consider is Sagittarius. In Section~\ref{sec:data}, we argued that $v_{los}\approx v_r$ for tracers with $\lvert \cos{\gamma} \rvert \geq 0.95$, but Sagittarius is relatively close-by at 16~kpc and has $\lvert \cos{\gamma} \rvert =0.93$, so one may question the inclusion of this object. However, once again we find  little change in the mass estimate without it, for all models. For example, the isotropic Hernquist model returned $M_{tot}=1.55\pm0.08\times10^{12}$\msun, which is almost identical to the result obtained using all the data (see Table~\ref{tab:results}). The scale radius is also unchanged within error ($17.0\pm2.2$\kpc).

We also investigate the effects on the mass estimate when tangential velocities are treated as parameters. To do this, we first obtain a mass estimate using only the Dinescu data (i.e. using only objects with complete velocity vectors), and find a slightly lower mass of $1.47\pm0.08 \times 10^{12}$ \msun. Next, we remove five tangential velocities from the data, and repeat the analysis treating those missing $v_t$'s as parameters. Repeating this process and removing five different $v_t$'s each time, we find that the $v_t$'s cannot be well estimated. However, treating $v_t$'s as parameters has little to no effect on the mass estimate, within error. There is one exception to the latter statement: when Pal 3's tangential velocity was removed, the mass estimate was reduced significantly. 

To investigate the influence of Pal 3 further, we performed an analysis using only the Dinescu data, but without Pal 3's $v_t$ value. Treating Pal 3's $v_t$ as a parameter, the mass estimate of the Milky Way fell by more than 50\% ($M_{tot}=0.76\pm0.06\times10^{12}$\msun). We also ran the analysis using only the Dinescu data, but without \emph{any} $v_t$'s. In this case, $M_{tot}=0.8 \pm 0.1\times10^{12}$\msun, which is similar to the estimate obtained in the former analysis. We note, however, that Pal 3 has the most uncertain $v_t$ measurement in the list (Table~\ref{tab:data}). It is evident that including measurement uncertainties in the analysis would reduce its leverage considerably.

Using all kinematic data, but removing Pal 3 from the analysis, also resulted in reduced mass estimates. Furthermore, the effect is observed regardless of the selected model (Table~\ref{tab:results}). Thus, Pal 3's proper motion, and indeed Pal 3 in general, has significant influence on the mass estimate of the Galaxy. This issue regarding Pal 3 confirms the finding of \cite{sakamoto2003}, who noted that high-velocity objects such as Pal 3 and Draco can have a significant effect on the mass estimate of the Galaxy.  As mentioned previously, we also test the effect of Draco on the mass estimate and find that it has little effect on the mass estimate (Table~\ref{tab:results}).

\begin{figure}
\centering
	\includegraphics[trim=0cm 0cm 0cm 0.2cm, clip=true, totalheight=0.39\textheight]{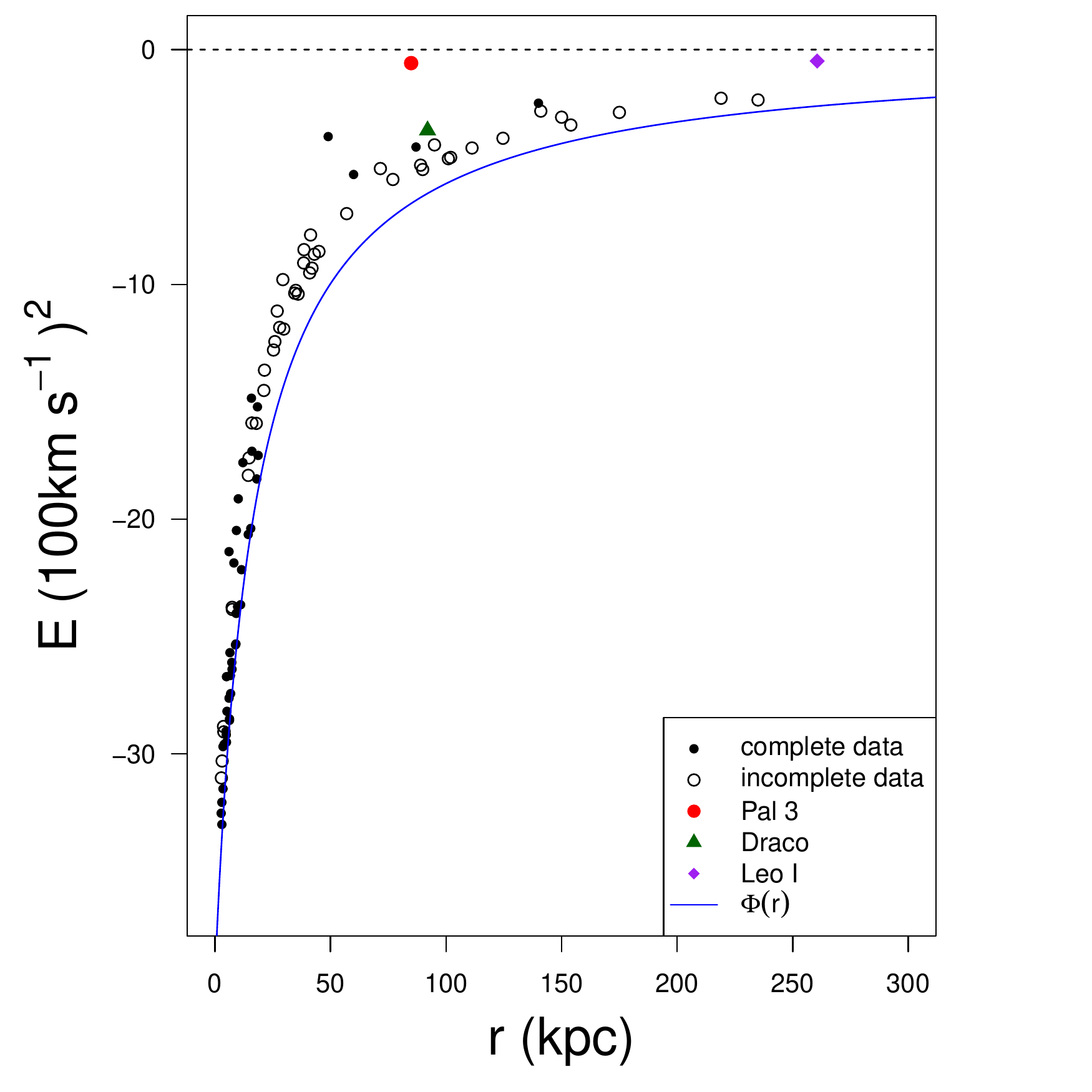}
	\caption[for table of contents]{Satellite energies given the model parameters from the isotropic Hernquist model fit. Satellites without tangential velocities are shown as open circles, and are plotted using the estimated $v_t$ value from the model fit. Unbound (escaping) objects would lie above the dotted line.}
	\label{fig:energies}
\end{figure}

To demonstrate the effectiveness of using $v_t$'s as parameters, consider Figure~\ref{fig:energies}. Using eq.~\ref{eq:scriptE} and the mean parameter values from the isotropic Hernquist model fit, we plot the negative of the tracers' specific energies as a function of $r$. Filled points are satellite data with complete velocity vectors, and hollow points are data with unknown $v_t$'s (plotted using the mean $v_t$ estimates from the Markov chain). As demonstrated by our simulations and tests with the Dinescu data, the $v_t$ values cannot be well estimated. However, the $v_t$ parameters appear to converge to an average $v_t$ for a given $r$ value, and this is reflected in the positions of the hollow points in Fig.~\ref{fig:energies}. 

The results demonstrate that in a small sample of data, some objects carry greater influence on the mass estimate than others. Furthermore, the variation in these results implies that it would be fruitful to weight the data by their measurement uncertainties. In a Bayesian analysis, however, a fully hierarchical approach is necessary to properly include the measurement uncertainties of the data, and a probability distribution for the errors must be assumed. We leave this analysis to a future paper, and instead perform three more approximate sensitivity analyses.

\begin{figure*}[t]
\centering
\includegraphics[trim=0cm 0cm 0cm 0cm, clip=true, totalheight=0.45\textheight]{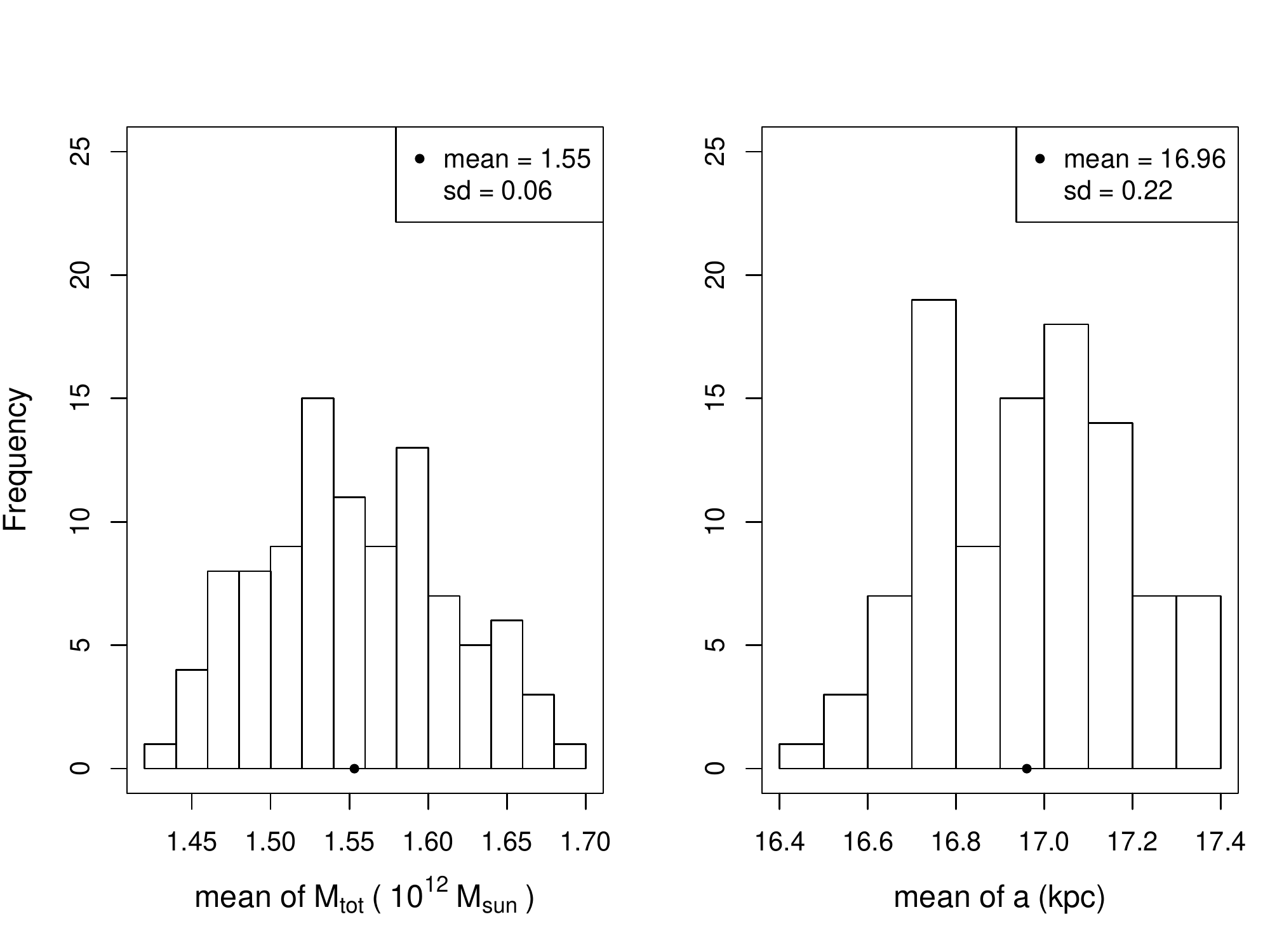} 
	\caption[]{Distribution of $M_{tot}$ and $a$ estimates from the third sensitivity analysis. The black dots show the mean of the estimates. The value of the mean and the standard deviation of the empirical distribution are shown in the legend.}
	\label{fig:sense}
\end{figure*}

The first two sensitivity analyses are extreme cases: 1) all the tangential velocities are increased by $2\Delta v_t$, and 2) all the tangential velocities are decreased by $2\Delta v_t$. In the third and more realistic sensitivity analysis, we randomly change each $v_t$ into a new tangential velocity $v_{t, new}$ via,
\begin{equation}\label{eq:sense}
 v_{t, new} = v_t + N(0, \Delta v_t) 
\end{equation}
where $N(0, \Delta v_t)$ represents a random number drawn from a normal distribution with mean zero and variance $\Delta v_t$. Using eq.~\ref{eq:sense}, we generate 100 data sets with new $v_t$ values, and then analyze these data sets assuming the isotropic Hernquist model. The estimates of $M_{tot}$ and $a$ from the 100 analyses have a distribution that confirms the results of the original analysis (Fig.~\ref{fig:sense}); the mean of the estimates is nearly identical to the result in Table~\ref{tab:results}.

The results of the sensitivity analyses show that a proper treatment of the measurement uncertainties is worth pursuing.  In future analyses, we plan to incorporate the measurement uncertainties of the data via a hierarchical model.

\section{Discussion and Future Prospects}\label{sec:discussion}

The results of this study are promising. Not only does the Bayesian analysis provide an effective way of incorporating complete and incomplete data, but it also enables easy calculation of probabilistic credible regions for the cumulative mass profile. Furthermore, even though this is a preliminary analysis, and mostly meant to lay the groundwork for future studies, our total mass estimates are similar to  other studies that use different methods.

Because our method returns a sample of parameter values representing the posterior distribution, it is easy to compare our results with mass estimates, at any radii, obtained in other studies. We can compute $M(r)$ credible regions from our Markov chain at any $r$ value, and obtain a mass estimate at that radius, with uncertainties.

A collection of total mass estimates within $r=100$kpc, from 10 different studies, has been compiled by \cite{courteau2014}. Our mass estimate at this radius assuming the isotropic Hernquist model is $M_{100}= 1.14 \times 10^{12}$\msun with a $95\%$ credible region of (1.05, 1.26)$\times 10^{12}$\msun, which is within the range of values listed in the review.

\cite{watkins2010} find the mass within 300 kpc to be $0.9\pm0.3\times10^{12}$\msun~for an isotropic model. Our estimate $M_{300}$ assuming an isotropic model is $1.39\times10^{12}$\msun with a 95\% credible region of (1.29, 1.53)$\times10^{12}$\msun. When they consider an anisotropic velocity distribution with $\beta$ derived from the observational data, however, they find $3.4\pm0.9\times10^{12}$\msun, in contrast to our $\beta=0.5$ constant anisotropic model that gives $M_{300}=1.35\times10^{12}$\msun, with a 95\% credible interval of (1.27,1.51)$\times10^{12}$\msun. 

\cite{2013deason} used BHB stars to trace the MW's mass, and found $M(r=50\text{kpc})$ to be approximately $4\times10^{11}$\msun, assuming a model of constant anisotropy with $\beta=0.5$. However, our mass estimate for the MW at 50kpc, using the Hernquist constant anisotropic model, is substantially higher at $9.5\times10^{11}$\msun~ with a 95\% credible interval of $(8.5,11.0)\times10^{11}$\msun. Even removing Pal 3 from the data set does not lower this estimate significantly, reducing it to $8.5\times10^{11}$\msun with a 95\% credible interval of $(7.5,9.6)\times10^{11}$\msun.

Using a truncated, flat rotation curve model, \cite{2005battaglia} found the mass of the MW dark matter halo to be $1.2^{+1.8}_{-0.5}\times10^{12}$\msun~, and with an NFW model found a virial radius of $0.8^{+1.2}_{-0.2}\times10^{12}$\msun. \cite{boylan2013} estimate the MW virial mass at $1.6\times10^{12}$\msun, with a 90\% confidence interval of 1.0 to $2.4\times10^{12}$\msun. \cite{sohn2013} use the timing argument of Leo I to arrive at a virial mass estimate $M_{vir}=3.15^{+1.58}_{-1.36}\times10^{12}$\msun. \cite{li2008} found the virial mass to be $2.4\times10^{12}$\msun, with a lower 95\% confidence level of $0.8\times10^{12}$\msun. Thus, our preliminary results are on par with many other studies that use different methodologies.

To our knowledge, no other studies besides \cite{sakamoto2003} have found Pal 3 to carry so much weight in the analysis. Pal 3's proper motion is already known (though with large uncertainties) and the satellite does not lie as far from the Galactic center as Leo I and other satellite dwarfs, which may have allowed its effect to go unnoticed.  Removing Pal 3's true $v_t$ from the analysis and treating it as a parameter lowered the mass significantly, suggesting that the tangential velocity is in the tail of the $v_t$ distribution at $r=84$\kpc. The dwarf galaxies in our data set, on the other hand, seem to have little individual influence on the mass estimate of the Galaxy even though some  have high velocities and large distances from the galactic center. 

Many improvements, challenges, and exploratory analyses remain:

\begin{enumerate}

\item{ One way to substantially improve the analysis is to incorporate measurement uncertainties via a hierarchical model, rather than the preliminary sensitivity analysis performed here. In the Bayesian paradigm, a probability density function of the measurement errors must be assumed. For example, for each data point the known measurement uncertainty may be used to define the variance of a Gaussian distribution centered on the measurement value. Objects with high influence when measurement errors are ignored (e.g.~Pal 3) might have reduced influence when measurement errors are included.}

\item{ An immediate challenge is finding models with distribution functions that are tractable and that describe the Milky Way in a more sophisticated manner. Using the NFW model is of particular interest because it is known to fit the dark matter halos of galaxies on many different scales, as well as groups and clusters of galaxies. The NFW DF, however, is not analytic. Although numerical solutions for the NFW DF have been derived, applying them in the Bayesian framework is more difficult because of the model's infinite mass. When a model's DF does not integrate to a finite mass (eq.~\ref{eq:fintM}) then it is not a proper probability distribution--- a requirement when applying Bayes' theorem. We are currently exploring the problem through an Approximate Bayesian Computation (ABC) algorithm, which allows for calculations of posterior distributions without explicit calculation of the likelihood.}

\item{ The DFs for all of the models employed here are analytic. Furthermore, the models are self-consistent--- i.e. we implicitly assume that the dark matter and the satellites follow the same distribution. However, it is possible that the tracers (GCs and dwarf galaxies) have a different distribution than the dark matter halo particles. For example, the tracers may have a Hernquist-type density profile $\rho_{\text{H}}(r)$ given by eq.~\ref{eq:rhor}, but may reside in an NFW gravitational potential given by
\begin{equation}\label{eq:NFW}
	 \Phi_{\text{NFW}}(r) =  -4\pi\rho_or^2_s \frac{\ln\left({1+r/r_s}\right)}{r/r_s}.
\end{equation}
In this situation, there are two extra parameters, $r_s$ and $\rho_o$, which correspond to the scale length and density parameter of the dark matter halo. We can derive the DF for such a model via the Eddington formula:
\begin{equation}\label{eq:Edd}
\begin{split}
	f(\mathcal{E}) &= \frac{1}{\sqrt{8}\pi^2}\int^{\mathcal{E}}_{0}{ \frac{1}{\sqrt{\mathcal{E}-\Psi}}\left(\frac{d^2\rho}{d\Psi^2}\right)}d\Psi \\
	& + \frac{1}{\sqrt{\mathcal{E}}}\left(\frac{d\rho}{d\Psi}\right)_{\Psi=0}
\end{split}
\end{equation}
where $\Psi=\Phi-\Phi_o$ is the relative potential \citep[see][]{binney2008}. For the Hernquist model, $\rho$ can be written as an analytic function of $\Psi$, and the integral can be evaluated in closed form.  For the case at hand, however, the relation between $\rho_{\text{H}}$ and $\Psi_{\text{NFW}}$ is a transcendental equation. Nevertheless, the integral required in the Eddington formula can be evaluated numerically. \cite{widrow2008}, for example, used this method to derive DFs for their self-consistent disk-bulge-halo galaxy models. A numerically derived DF may still be used with our method, as long as it is a normalized probability distribution. We plan to implement models of this type in future studies of tracer populations.}

\item{ Different velocity anisotropy formalisms are also of interest. For example, other Hernquist model DFs of different anisotropies are discussed by \cite{1991cuddeford} and \cite{baes2002}. The former derived a velocity anisotropy that is a generalization of the OM-type anisotropy, where another parameter $\beta_0$ is introduced such that
\begin{equation}\label{eq:beta0}
 \beta(r) = \frac{r^2 + \beta_0r_a^2}{r^2 + r_a^2}.
\end{equation}
When $\beta_0 = 0$, eq.~\ref{eq:beta0} reduces to OM-type anisotropy. As $r_a\rightarrow\infty$, $\beta(r)\rightarrow\beta_0$, in constrast to eq.~\ref{eq:OM}. \cite{baes2002} derive a Hernquist DF using this formalism, and show that only four values of $\beta_0$ lead to DFs that can be expressed in terms of elementary functions. The simplest of these DFs occurs when $\beta_0=0.5$, while the other DFs are "...somewhat more elaborate" and not provided~\citep{baes2002}. }

\item{We will explore biases that may occur due to selection effects. In the Hernquist simulations used here, some tracers are unrealistically far from the Galactic center (e.g. more than 500kpc away), while our kinematic data set has a range from $r=3$kpc to $261$ kpc. Imposing a more realistic range on simulated data may or may not introduce biases in parameter estimates.}

\item{Further along the line, it will be exciting to apply the method presented here to large datasets of field halo stars, leading up to the GAIA data. \cite{sakamoto2003} showed that including many field horizontal branch stars greatly reduced the effect of high-velocity objects (such as Draco and Pal 3) on the mass estimate of the Milky Way. Therefore, it can be expected that the accurate and abundant kinematic data from GAIA will also improve our mass estimates in a major way and decrease the effect of outliers.}

\item{The method outlined in this paper could also be extended to obtain mass estimates of other galaxies for which tracer objects will have only the projected positions and line of sight velocities.}

\end{enumerate}

\section*{Summary}

We have introduced a method to estimate the mass of the Milky Way that incorporates both complete and incomplete data for positions and velocities of tracers. The method treats unknown tangential velocities as parameters in the model. Simulations showed that although the tangential velocities cannot be well constrained, treating $v_t$'s as parameters has little effect on the mass estimate when other complete velocity vectors are available. An exception does occur, however, when a tracer has an unusually extreme tangential velocity (e.g.~Pal 3).

Under simple assumptions of a Hernquist-like halo potential and modest anisotropy, we find $M_{tot}\approx1.3-1.5\times10^{12}$\msun, in good agreement with other recent work. For an isotropic model we find $M_{tot}=1.55\times10^{12}$\msun~ with a 95\% credible interval of $(1.42,1.73)\times10^{12}$\msun, and a scale radius of $a=16.9$kpc. We also report the mass contained within 260kpc: $1.37\times10^{12}$\msun, with a 95\% credible interval of $(1.27,1.51)\times10^{12}$\msun.

In future research, we will be incorporating measurement uncertainties into the analysis and will test more extensively for parameter biases. We also plan to use the NFW model and find other DF's to use in the GME code. The method outlined here will eventually be applied to extragalactic studies, where the complete velocity vectors of the tracers are unknown.

\section*{Acknowledgements}

WEH  and LMW acknowledge the financial support of NSERC. GME thanks the reviewer for very useful comments and suggestions, and would also like to thank Aaron Springford for useful discussions about sensitivity analyses.

\bibliographystyle{apj}

\bibliography{myrefs}

\label{lastpage}

\end{document}